\documentclass[12pt]{article}

\usepackage{amsfonts,amssymb,graphicx,a4wide}
\usepackage{amsmath}

\begin{document}

\begin{titlepage} 

\begin{centering} 

{\large {\bf Liouville Decoherence in a Model of Flavour Oscillations 
in the presence of Dark Energy}}

\vspace{0.5cm}
{\bf Nick Mavromatos and Sarben Sarkar} \\
\vspace{0.5cm}
{\it Department of Physics\\
King's College London\\
Strand, London WC2R 2LS,UK}

\end{centering} 

\vspace{0.5cm}
\begin{abstract}

We study in some detail the master equation, and its solution
in a simplified case modelling flavour oscillations
of a two-level system, 
stemming from the Liouville-string approach to quantum space time foam.
In this framework we discuss the appearance of diffusion terms and 
decoherence due to the interaction of low-energy string matter with
space-time defects, such as D-particles 
in the specific model of ``D-particle foam'', as well as 
dark energy contributions. We pay particular attention to 
contrasting  
the decoherent role of a cosmological constant
in inducing exponential quantum damping 
in the evolution of low-energy observables,
such as the probability of flavour oscillations, 
with the situation where the dark energy 
relaxes to zero for asymptotically large times, 
in which case such a damping is absent. Our findings may be of interest 
to (astrophysical) tests of quantum space-time foam models in the 
not-so-distant future.

\end{abstract}

\end{titlepage}

\section{Introduction and Motivation}

In recent years there has been a debate on whether microscopic black holes
can induce quantum decoherence at a microscopic level. The presence of
quantum-fluctuating microscopic horizons, of radius of the order of Planck
length ($10^{-35}$ m), may give space-time a ``foamy'' structure, causing
decoherence of matter propagating in it. In particular, it has been
suggested~\cite{hawking} that such Planck-scale black holes and other
topological fluctuations in the space-time background cause a breakdown of
the conventional S-matrix description of asymptotic scattering in local
quantum field theory. This may lead to experimentally testable predictions,
at least in principle, for instance as regards the so-called sensitive
particle-physics probes of quantum mechanics~\cite{ehns,kaons,cplear}.

It must be pointed out that this suggestion is invalidated if there is \emph{%
holography} in quantum gravity~\cite{thooft}, such that any information of
quantum numbers of matter, that at first sight appears to be lost into the
horizon, is somehow reflected back on the horizon surface, thereby
maintaining quantum coherence. This may happen, for instance, in some highly
supersymmetric effective theories of strings \cite{Maldacena}, which however
do not represent realistic low-energy theories of quantum gravity.
Supersymmetry breaking complicates the issue, spoiling complete holography.
Recently, S. Hawking, inspired by the above recent ideas in string theory,
has also argued against the loss of coherence in a \emph{Euclidean} quantum
theory of gravity. In such a model, summation over trivial and non-trivial
(black-hole) space-time topologies in the path over histories makes an
asymptotic observer ``unsure'' as to the existence of the microscopic black
hole fluctuation thus resulting in no loss of quantum coherence. However,
this sort of argument is plagued not only by the Euclidean formalism, with
its concomitant problems of analytic continuation, but also by a lack of a
concrete rigorous proof, at least up to now.

Therefore, for our purposes we still consider the matter of
quantum-gravity-induced decoherence as wide open and worthy of further
phenomenological exploitation. This is the point of view we shall take in
this work. We shall restrict ourselves to a specific framework for analyzing
decoherent propagation of low-energy matter in foamy space-time backgrounds
in the context of string theory~\cite{strings,polch}, the so-called
Liouville-string~\cite{ddk} decoherence~\cite{emn}. Our motivation for using
string theory is that it appears at present to be the best controlled theory
of quantum gravity available to date. In this respect we also mention that
there are other interesting approaches to quantum space-time foam, which
also lead to experimental predictions, for instance the ``thermal bath''
approach advocated in \cite{garay}, according to which the foamy
gravitational environment may behave as a thermal bath, inducing decoherence
and diffusion in the propagating matter, as well as quantum damping in the
evolution of low-energy observables, features which are, at least in
principle, testable experimentally. As we shall see later on, similar
behaviour is exhibited by the specific models of foam that we study here,
which may characterize modern versions of string theory~\cite{polch},
specifically the D-particle foam model of \cite{emw,recoil}, based on
point-like membrane defects in space time (D-particles).

In the presence of decoherence the S-matrix of the effective low-energy
field theory would then have to be replaced by a linear non-factorisable
superscattering operator $\not{S}$ relating initial and final-state density
matrices~\cite{hawking} 
\begin{equation}
\rho _{out}=\not{S}\rho _{in}.
\end{equation}%
If this is correct then the usual formulation of quantum mechanics has to be
modified. Arguments have been put forward for this modification of the
Liouville equation to take the form~\cite{ehns}%
\begin{equation}
\partial _{t}\rho =\frac{i}{\hbar }\left[ \rho ,H\right] +\not{\delta}H\rho .
\label{liouvqm}
\end{equation}%
Equations of this form are encountered in the description of the time
evolution of the state of open quantum mechanical systems where $\not{\delta}%
H\rho $ has a Lindblad form~\cite{lindblad}. In such systems observable
degrees of freedom are coupled to unobservable components which are
effectively integrated over. Initial pure states evolve into mixed ones and
so 
\begin{equation}
\not{S}\neq SS^{\dag }
\end{equation}%
where $S=e^{iHt}$. In these circumstances Wald~\cite{wald} has shown that
CPT is violated, at least in its strong form, i.e. there is \textit{no}
unitary invertible operator $\Theta $ such that 
\begin{equation}
\Theta \overline{\rho }_{\mathrm{in}}=\rho _{\mathrm{out}}\mathrm{.}
\end{equation}

Such considerations have more recently again come to the fore because of
current neutrino data including LSND data~\cite{reviewnu}. There is
experimental evidence, that the neutrino has mass which leads to neutrino
oscillations. However LSND results appear consistent with the dominance of
anti-neutrino oscillations $\overline{\nu }_{e}\rightleftarrows $ $\overline{%
\nu }_{\mu }$ over neutrino oscillations. In particular, provided LSND
results turn out to be correct, which is unclear at present, there is
evidence for CPT violation. It has been suggested recently~\cite{barenboim}
that Planck scale quantum decoherence may be a relevant contribution to the
CPT violation seen in the experiments of LSND. Other examples of flavour
oscillating systems with quite different mass scales are furnished by $B%
\overline{B}$ and $K\overline{K}$ systems~\cite{kaons}. The former, because
of the large masses involved, provides a particularly sensitive system for
investigating the Planck scale fluctuations embodied by space-time foam. In
all these cases experiments, such as CPLEAR~\cite{cplear}, provide very low
bounds on CPT violation which are not inconsistent with dimensional analysis
estimates for the magnitudes of effects from space-time foam. These systems
have been analyzed within a dynamical semigroup approach to quantum Markov
processes. Once the framework has been accepted then a master equation for
finite-dimensional systems ensued which can be characterized by a small set
of parameters. This approach is somewhat phenomenological and is primarily
used to fit data. Consequently it is important to obtain a better
understanding of the nature of decoherence from a more fundamental viewpoint.

An additional, and perhaps more plausible~\cite{mavcptdecoh}, reason for
considering quantum decoherence models of quantum gravity, comes from recent
astrophysical evidence on a current-era acceleration of our Universe.
Indeed, observations of distant supernovae~\cite{snIa}, as well as WMAP data~%
\cite{wmap} on the thermal fluctuations of the cosmic microwave background
(CMB), indicate that our Universe is at present in an accelerating phase,
and that 73\% of its energy-density budget consists of an unknown substance,
termed \emph{Dark Energy}. Best-fit models of such data include
Einstein-Friedman-Robertson-Walker Universes with a non-zero cosmological 
\emph{constant}. However, the data are also currently compatible with
(cosmic) time-dependent vacuum-energy-density components, relaxing
asymptotically to zero~\cite{steinhardt}. In colliding brane-world models
the dark energy component of the Universe is considered to be a
non-equilibrium energy density of the (observable) brane world~\cite%
{gravanis,emw}. This density is identified with the central charge surplus
of the supercritical $\sigma $-models describing the (recoil) string
excitations of the brane after the collision.The relaxation of the
dark-energy density component can be a purely stringy feature of the
logarithmic conformal field theory~\cite{lcftmav} describing the D-brane
recoil~\cite{kmw} in a (perturbative) $\sigma $-model framework. We shall
discuss below how the theory of \emph{non-critical} strings deals with the
twin problems of cosmological constant, and dark energy. It is worth
pointing out that the non-critical (Liouville) string~\cite{ddk} provides a
rather unified formalism for dealing not only with Universes with a non-zero
cosmological constant in string theory, but in general with decoherent
quantum space-time foam backgrounds, that include microscopic
quantum-fluctuating black holes~\cite{emn}.

\textit{\ }

\section{Liouville decoherence: general formalism}

Given the very limited understanding of gravity at the quantum level, the
analysis of modifications of the quantum Liouville equation implied by
non-critical strings can only be approximate and should be regarded as
circumstantial evidence in favour of the dissipative master equation. In the
context of two-dimensional toy black holes\cite{2dbhstring} and in the
presence of singular space-time fluctuations there are believed to be
inherently unobservable delocalised modes which fail to decouple from light
(i.e. the observed) states. The effective theory of the light states which
are measured by local scattering experiments can be described by a
non-critical Liouville string. This results in an irreversible temporal
evolution in target space with decoherence and associated entropy production.

The following master equation for the evolution of stringy low-energy matter
in a non-conformal $\sigma $-model~can be derived\cite{emn} 
\begin{equation}
\partial _{t}\rho =i\left[ \rho ,H\right] +:\beta ^{i}\mathcal{G}_{ij}\left[
g^{j},\rho \right] :  \label{master}
\end{equation}%
where $t$ denotes time (Liouville zero mode), the $H$ is the effective
low-energy matter Hamiltonian, $g^{i}$ are the quantum background target
space fields, $\beta ^{i}$ are the corresponding renormalization group $%
\beta $ functions for scaling under Liouville dressings and $\mathcal{G}%
_{ij} $ is the Zamolodchikov metric \cite{zam,kutasov} in the moduli space
of the string. The double colon symbol in (\ref{master}) represents the
operator ordering $:AB:=\left[ A,B\right] $ . The index $i$ labels the
different background fields as well as space-time. Hence the summation over $%
i,j$ in (\ref{master}) corresponds to a discrete summation as well as a
covariant integration $\int d^{D+1}y\,\sqrt{-g}$\bigskip\ where $y$ denotes
a set of $\left( D+1\right) $-dimensional target space-time co-ordinates and 
$D$ is the space-time dimensionality of the original non-critical string.

\subsection{D-particle Foam and Master Equation}

The discovery of new solitonic structures in superstring theory~\cite{polch}
has dramatically changed the understanding of target space structure. These
new non-perturbative objects are known as D-branes and their inclusion leads
to a scattering picture of space-time fluctuations. The study of D-brane
dynamics has been made possible by Polchinski's realization~\cite{polch}
that such solitonic string backgrounds can be described in a conformally
invariant way in terms of world sheets with boundaries. On these boundaries
Dirichlet boundary conditions for the collective target-space coordinates of
the soliton are imposed. Heuristically, when low energy matter given by a
closed (or open) string propagating in a $\left( D+1\right) $-dimensional
space-time collides with a very massive D-particle embedded in this
space-time, the D-particle recoils as a result. Since there are no rigid
bodies in general relativity the recoil fluctuations of the brane and their
effectively stochastic back-reaction on space-time cannot be neglected.

Based on these considerations, a model for a supersymmetric space-time foam
has been suggested in \cite{emw} . The model is based on parallel brane
worlds (with three spatial large dimensions), moving in a bulk space time
which contains a ``gas'' of D-particles. The number of parallel branes used
is dictated by the requirements of target-space supersymmetry in the limit
of zero-velocity branes. One of these branes represents allegedly our
Observable Universe. As the brane moves in the bulk space, D-particles cross
the brane in a random way. From the point of view of an observer in the
brane the crossing D-particles will appear as space-time defects which
flashing on and off , i.e. microscopic space-time fluctuations. This will
give the four-dimensional brane world a ``D-foamy'' structure.

\begin{figure}[tb]
\begin{centering} 

\includegraphics[width=5cm]{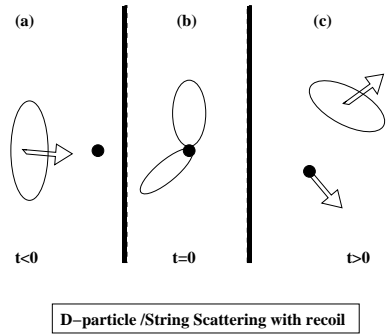}

\end{centering} 
\caption{\textit{Schematic picture of the scattering of a string matter
state on a D-particle, including recoil of the latter. The sudden impulse at 
$t=0$, implies a back reaction onto the space time, which is described by a
logarithmic conformal field theory. The method allows for the perturbative
calculation of the induced space-time distortion due to the entangled state
in (b).}}
\label{fig:dfoam}
\end{figure}
Closed and open strings propagate on the brane. Each time these strings
cross with a D-particle, there is a possibility of being attached to it, as
indicated in Fig. \ref{fig:dfoam}. The entangled state causes a back
reaction onto the space-time, which can be calculated perturbatively using
the formalism of logarithmic conformal field theory ~\cite{kmw}. Details are
reviewed in Appendix B.

Using this model for space-time fluctuations we will obtain an expression
for the induced space-time distortion as a result of D-particle recoil. The
set $\left\{ g^{i}\right\} $ includes the graviton fields $g_{MN}$ where $M$
and $N$ are target space-time indices and in the weakly coupled string limit
we can show that: 
\begin{equation}
g_{\ell m}=\delta _{\ell m},\,g_{00}=-1,g_{0\ell}=\varepsilon \left( \varepsilon
y_{\ell}+u_{\ell}t\right) \Theta _{\varepsilon }\left( t\right) ,\; \ell, m = 
1, \ldots ,D
\label{recmetr}
\end{equation}%
where the suffix $0$ denotes temporal (Liouville) components and 
\begin{eqnarray}
\Theta _{\varepsilon }\left( t\right) &=&\frac{1}{2\pi i}\int_{-\infty
}^{\infty }\frac{dq}{q-i\varepsilon }e^{iqt},  \label{heaviside} \\
u_{\ell} &=&\left( k_{1}-k_{2}\right) _{\ell}\;,  \notag
\end{eqnarray}%
with $k_{1}\left( k_{2}\right) $ the momentum of the propagating
closed-string state before (after) the recoil; $y_{n}$ are the spatial
collective coordinates of the D particle and $\varepsilon ^{-2}$ is
identified with the target Minkowski time $t$ for $t\gg 0$ after the
collision~\cite{kmw} (see Appendix B). These relations have been calculated
for non-relativistic branes where $u_{n}$ is small and require the machinery
of logarithmic conformal field theory. Now for large $t,$ to leading order,

\begin{equation}
g_{0\ell}\simeq \overline{u}_{\ell}\equiv \frac{u_{\ell}}{\varepsilon }\propto \frac{%
\Delta p_{\ell}}{M_{P}}  \label{recoil}
\end{equation}%
where $\Delta p_{\ell}$, $\ell = 1, \ldots, D$, is 
the momentum transfer during a collision and $M_{P}$
is the Planck mass (actually, to be more precise, $M_{P}=M_{s}/g_{s}$, where 
$g_{s}<1$ is the (weak) string coupling, and $M_{s}$ is a string mass
scale); so $g_{0i}$ is constant in space-time but depends on the energy
content of the low energy particle and $R_{MN}=0$ . Since we are interested
in fluctuations of the metric the indices $i$ will correspond to the pair $%
M,N$.

However, as already mentioned in the introduction, recent astrophysical
observations from different experiments all seem to indicate that $73$\% of
the energy of the Universe is in the form of dark energy. Best fit models
give the positive cosmological constant Einstein-Friedman Universe as a good
candidate to explain these observations. For such de Sitter backgrounds $%
R_{MN}\propto \Omega g_{MN}$ with $\Omega >0$ a cosmological constant. Also
in a perturbative derivative expansion (in powers of $\alpha ^{\prime }$
where $\alpha ^{\prime }=l_{s}^{2}$ is the Regge slope of the string and $%
l_{s}$ is the fundamental string length) in leading order 
\begin{equation}  \label{fixedpoint}
\beta _{\mu \nu }=\alpha ^{\prime }R_{\mu \nu }=\alpha ^{\prime }\Omega
g_{\mu \nu }
\end{equation}
and 
\begin{equation}
\mathcal{G}_{ij}=\delta _{ij}.
\end{equation}

This leads to 
\begin{equation}
\partial _{t}\rho =i\left[ \rho ,H\right] +\,\alpha ^{\prime }\Omega :g_{MN}%
\left[ g^{MN},\rho \right] :  \label{master2}
\end{equation}
For a weak-graviton expansion about flat space-time, $g_{MN}=\eta
_{MN}+h_{MN}$, and 
\begin{equation}
h_{0\ell}\simeq \overline{u}_{\ell}\equiv \frac{u_{\ell}}{\varepsilon }\propto \frac{%
\Delta p_{\ell}}{M_{P}}.  \label{recoil2}
\end{equation}
If an antisymmetric ordering prescription is used, then the master equation
for low energy string matter assumes the form%
\begin{equation}
\overset{.}{\partial _{t}\rho _{Matter}}=i\left[ \rho _{Matter},H\right]
-\,\Omega \left[ h_{0\ell},\left[ h^{0\ell},\rho _{Matter}\right] \right]
\label{master3}
\end{equation}%
( when $\alpha ^{\prime }$ is absorbed into $\Omega )$. In view of the
previous discussion this can be rewritten as%
\begin{equation}
\overset{.}{\partial _{t}\rho _{Matter}}=i\left[ \rho _{Matter},H\right]
-\,\Omega \left[ \overline{u}_{\ell},\left[ \overline{u}^{\ell},\rho _{Matter}%
\right] \right]~.  \label{master4}
\end{equation}
thereby giving the master equation for Liouville decoherence in the model of
a D-particle foam with a cosmological constant.

\section{Destruction of interference}

The master equation which has been derived is of relevance to the study of
general features of gravity induced decoherence.The above D-particle
inspired approach deals with possible non-perturbative quantum effects of
gravitational degrees of freedom. The analysis is totally unrelated to the
phenomenology of dynamical semigroups which does not embody specific
properties of gravity. Indeed the phenomenology is sufficiently generic that
other mechanisms of decoherence such as the MSW effect can be incorporated
within the same framework. Consequently an analysis which is less generic
and is related to the specific \ decoherence implied by non-critical strings
is necessary.

It is sufficient to study a massive non-relativistic particle propagating in
one dimension to establish qualitative features of D-particle decoherence.
The environment will be taken to consist of both gravitational and
non-gravitational degrees of freedom; hence we will consider a
generalisation of quantum Brownian motion for a particle which has
additional interactions with D-particles. This will allow us to compare
qualitatively the decoherence due to different environments.The
non-gravitational degrees of freedom in the environment (in a thermal state)
are conventionally modelled by a collection of harmonic oscillators with
masses $m_{n}$, frequency $\omega _{n}$ and co-ordinate operator $\widehat{q}%
_{n}$ coupled to the particle co-ordinate $\widehat{x}$ by an interaction of
the form $\sum_{n}g_{n}\widehat{x}\widehat{q}_{n}$. The master equation
which is derived can have time dependent coefficients due to the competing
timescales, e.g. relaxation rate due to coupling to the thermal bath, the
ratio of the time scale of the harmonic oscillator to the thermal time scale
etc. However an ab initio calculation of the time-dependence is difficult to
do in a rigorous manner. It is customary to characterise the
non-gravitational environment by means of its spectral density $I\left(
\omega \right) \left( =\sum_{n}\delta \left( \omega -\omega _{n}\right) 
\frac{g_{n}^{2}}{2m_{n}\omega _{n}}\right) $. The existence of the different
time scales leads in general to non-trivial time dependences in the
coefficients in the master equation which are difficult to calculate in a
rigorous manner\cite{BLH1992}. The dissipative term in (\ref{master4})
involves the momentum transfer operator due to recoil of the particle from
collisions with D-particles (\ref{recoil}). This transfer process will be
modelled by a classical Gaussian random variable $r$ which multiplies the
momentum operator $\widehat{p}$ for the particle: 
\begin{equation}
\overline{u_{x}}\qquad \rightarrow \qquad \frac{r}{M_{P}}\widehat{p}
\label{trnsf}
\end{equation}%
Moreover the mean and variance of $r$ are given by 
\begin{equation}
\left\langle r\right\rangle =0~,\qquad \mathrm{and}\qquad \left\langle
r^{2}\right\rangle =\sigma ^{2}~.  \label{random}
\end{equation}
On amalgamating the effects of the thermal and D-particle environments, we
have for the reduced master equation for the matter (particle) density
matrix $\rho $ (on dropping the Matter index)%
\begin{equation}
i\frac{\partial }{\partial t}\rho =\frac{1}{2m}\left[ \widehat{p}^{2},\rho %
\right] -i\Lambda \left[ \widehat{x},\left[ \widehat{x},\rho \right] \right]
+\frac{\gamma }{2}\left[ \widehat{x},\left\{ \widehat{p},\rho \right\} %
\right] -i\Omega r^{2}\left[ \widehat{p},\left[ \widehat{p},\rho \right] %
\right]  \label{master5}
\end{equation}%
where $\Lambda ,\gamma $ and $\Omega $ are real time-dependent coefficients.
As discussed in Appendix B (\ref{omegatotal}) a possible model for $\Omega
\left( t\right) $ is 
\begin{equation}
\Omega \left( t\right) =\Omega _{0}+\frac{\widetilde{\gamma }}{a+t}+\frac{%
\widetilde{\Gamma }}{1+bt^{2}}  \label{cosmological}
\end{equation}%
where $\Omega _{0}$, $\widetilde{\gamma }$, $a$, $\widetilde{\Gamma }$ and $%
b $ are positive constants, including appropriate powers of $\alpha ' $. 
The quantity $\widetilde{\gamma }<1$ contains
information on the density of D-particle defects on a four-dimensional
world. From the earlier derivation of the master equation it is clear that
the last term of the r.h.s. of (\ref{cosmological}) represents a
time-dependent cosmological constant contribution. The form of this time
dependence, just as other issues to do with the cosmological constant, is a
matter of debate. Arguments, for example, arising from quintessence~\cite%
{quintess} and cosmon scalar fields lead to a behaviour which is compatible
with (\ref{cosmological}). Similar behaviour is also obtained in dark.energy
relaxation models in the context of non-critical strings (linear-dilaton
models~\cite{aben,emngrf} or colliding-brane worlds (recoil)~\cite%
{gravanis,emw}). The time dependence of $\gamma $ and $\Lambda $ can be
calculated in the weak coupling limit for general $n$ (i.e. ohmic, $n=1$ and
non-ohmic $n\neq 1$ environments)$\ $where 
\begin{equation}
I\left( \omega \right) =\frac{2}{\pi }m\gamma _{0}\omega \left[ \frac{\omega 
}{\varpi }\right] ^{n-1}e^{-\omega ^{2}/\varpi ^{2}}  \label{spectral}
\end{equation}%
and $\varpi $ is a cut-off frequency. The precise time dependence is
governed by $\Lambda \left( t\right) =\int_{0}^{t}ds\,\nu \left( s\right) $
and $\gamma \left( t\right) =\int_{0}^{t}ds\,\nu \left( s\right) s$ where $%
\nu \left( s\right) =\int_{0}^{\infty }d\omega \,I\left( \omega \right)
\coth \left( \beta \hbar \omega /2\right) \cos \left( \omega s\right) $. For
the ohmic case, in the limit $\hbar \varpi \ll k_{B}T$ followed by $\varpi
\rightarrow \infty $, $\Lambda $ and $\gamma $ are given by $m\gamma
_{0}k_{B}T$ and $\gamma _{0}$ respectively after a rapid initial transient.
For high temperatures $\Lambda $ and $\gamma $ have a powerlaw increase with 
$t$ for the subohmic case whereas there is a rapid decrease in the
supraohmic case. However for our considerations it is adequate to restrict
attention to the ohmic high temperature limit~\cite{WGU1989}.

\section{Solution of the master equation}

The master equation of (\ref{master5}) can be solved. It is useful to
introduce the operator 
\begin{equation}
\widehat{D}=\exp \left( i\left( k\widehat{x}+\Delta \widehat{p}\right)
\right)  \label{Ddef}
\end{equation}%
and the transform 
\begin{equation}
\widetilde{\rho }\left( \kappa ,\Delta \right) =tr\left( \rho \exp \left(
i\left( \kappa \widehat{x}+\Delta \widehat{p}\right) \right) \right) .
\label{transform}
\end{equation}%
The master equations then takes the form%
\begin{equation}
\frac{\partial }{\partial t}\widetilde{\rho }+\left( -\frac{\kappa }{m}%
+\gamma \Delta \right) \frac{\partial }{\partial \Delta }\widetilde{\rho }%
=-\Lambda \Delta ^{2}\widetilde{\rho }-\Omega \left( t\right) r^{2}\kappa
^{2}\widetilde{\rho }.  \label{mastertransform}
\end{equation}%
This is solved by~\cite{CMS1985} 
\begin{equation}
\widetilde{\rho }\left( \kappa ,\Delta ,t\right) =\digamma \left( e^{-\gamma
t}\left( \Delta -\frac{\kappa }{m\gamma }\right) ,\kappa \right) \mathcal{P}%
_{I}\left( \kappa ,\Delta ,t\right)  \label{ansatz2}
\end{equation}%
where $\mathcal{P}_{I}$ is a particular solution of \ref{mastertransform}
and $\digamma \left( e^{-\gamma t}\left( \Delta -\frac{\kappa }{m\gamma }%
\right) ,\kappa \right) $ is an arbitrary solution of 
\begin{equation}
\frac{\partial }{\partial t}\widetilde{\rho }+\left( -\frac{\kappa }{m}%
+\gamma \Delta \right) \frac{\partial }{\partial \Delta }\widetilde{\rho }=0.
\label{DerivativePart}
\end{equation}%
for an arbitrary $\digamma $.

We can write 
\begin{equation}
\mathcal{P}_{I}\left( \kappa ,\Delta ,t\right) =\exp \left( f_{1}\left(
t\right) \kappa ^{2}+f_{2}\left( t\right) \Delta ^{2}+f_{3}\left( t\right)
\kappa \Delta \right) .  \label{ansatz3}
\end{equation}%
It is then necessary that 
\begin{eqnarray}
\frac{d}{dt}f_{1}-\frac{f_{3}}{m} &=&-r^{2}\Omega \left( t\right)
\label{particular1} \\
\frac{d}{dt}f_{2}+2\gamma f_{2} &=&-\Lambda  \label{particular2} \\
\frac{d}{dt}f_{3}-\frac{2}{m}f_{2}+\gamma f_{3} &=&0.  \label{particular3}
\end{eqnarray}%
Now (\ref{particular1}) \ is a stochastic differential equation.We can
consider $r\left( t\right) $to be a (finite variance) Gaussian variable with
variance $\sigma ^{2}$ and vanishing mean. In the ohmic high temperature
limit and in the mean we can solve these coupled equations, for initial
conditions $f_{1}\left( 0\right) =f_{2}\left( 0\right) =f_{3}\left( 0\right)
=0$ to give 
\begin{eqnarray}
f_{1}\left( t\right) &=&\frac{1}{2m\gamma ^{2}\sqrt{b}}\left( k_{B\;}T\sqrt{b%
}\left( -2\gamma t-\left( 1-e^{-2\gamma t}\right) +4\left( 1-e^{-\gamma
t}\right) \right) -\right.  \notag \\
&~&\left. 2m\sigma ^{2}\gamma ^{2}\widetilde{\Gamma }\tan ^{-1}\left( \sqrt{b%
}t\right) \right)  \notag \\
&~&  \label{soln1} \\
f_{2}\left( t\right) &=&\frac{1}{2}mk_{B}T\left( e^{-2\gamma t}-1\right)
\label{soln2} \\
f_{3}\left( t\right) &=&\frac{k_{B}T}{\gamma }\left( 2\left( e^{-\gamma
t}-1\right) +\left( 1-e^{-2\gamma t}\right) \right) .  \label{soln3}
\end{eqnarray}%
on taking $\Omega _{0}=\widetilde{\gamma }=0$. In order to study
interference we need to understand the behaviour of $\rho \left( x,x^{\prime
}\right) $. $\rho $ and $\widetilde{\rho }$ are related by 
\begin{equation}
\rho \left( x,x^{\prime },t\right) =\frac{1}{2\pi }\int_{-\infty }^{\infty
}d\kappa e^{-i\kappa \left( x+x^{\prime }\right) /2}\widetilde{\rho }\left(
\kappa ,x-x^{\prime },t\right) .  \label{transform2}
\end{equation}

For an initial state which is a linear superposition $\left|
k_{1}\right\rangle +\left| k_{2}\right\rangle $ of momentum eigenstates 
\begin{equation}
\widetilde{\rho }\left( \kappa ,\Delta ,t=0\right) =\pi e^{i\kappa \Delta
/2}\left( \delta \left( \kappa \right) \left( e^{i\Delta k_{1}}+e^{i\Delta
k_{2}}\right) +\delta \left( \kappa +k_{2}-k_{1}\right) e^{i\Delta
k_{2}}+\delta \left( \kappa +k_{1}-k_{2}\right) e^{i\Delta k_{1}}\right)
\label{initial}
\end{equation}%
and so 
\begin{equation}
\digamma \left( \Delta ,\kappa \right) =\pi e^{i\frac{\kappa }{2}\left(
\Delta +\frac{\kappa }{m\gamma }\right) }\left( 
\begin{array}{c}
\delta \left( \kappa \right) \left( e^{i\left( \Delta +\frac{\kappa }{%
m\gamma }\right) k_{1}}+e^{i\left( \Delta +\frac{\kappa }{m\gamma }\right)
k_{2}}\right) +\delta \left( \kappa +k_{2}-k_{1}\right) e^{^{i\left( \Delta +%
\frac{\kappa }{m\gamma }\right) k_{2}}} \\ 
+\delta \left( \kappa +k_{1}-k_{2}\right) e^{i\left( \Delta +\frac{\kappa }{%
m\gamma }\right) k_{1}}%
\end{array}%
\right) .  \label{initial2}
\end{equation}%
This implies that 
\begin{equation}
\rho \left( x,x-\Delta ,t\right) =\frac{1}{2}\left( 
\begin{array}{c}
e^{f_{2}\left( t\right) \Delta ^{2}+ik_{1}g\left( t,0,\Delta \right)
}+e^{f_{2}\left( t\right) \Delta ^{2}+ik_{2}g\left( t,0,\Delta \right) }+ \\ 
e^{f_{1}\left( t\right) \left( k_{1}-k_{2}\right) ^{2}+f_{2}\left( t\right)
\Delta ^{2}+f_{3}\left( t\right) \left( k_{1}-k_{2}\right) \Delta }e^{i\left[
\left( k_{1}-k_{2}\right) \left( \frac{\Delta }{2}-x\right) +\frac{\left(
k_{1}+k_{2}\right) }{2}g\left( t,k_{1}-k_{2},\Delta \right) \right] }+ \\ 
e^{f_{1}\left( t\right) \left( k_{2}-k_{1}\right) ^{2}+f_{2}\left( t\right)
\Delta ^{2}+f_{3}\left( t\right) \left( k_{2}-k_{1}\right) \Delta }e^{i\left[
-\left( k_{1}-k_{2}\right) \left( \frac{\Delta }{2}-x\right) +\frac{\left(
k_{1}+k_{2}\right) }{2}g\left( t,k_{2}-k_{1},\Delta \right) \right] }%
\end{array}%
\right)  \label{interference}
\end{equation}%
where $g\left( t,\kappa ,\Delta \right) =e^{-\gamma t}\left( \Delta -\frac{%
\kappa }{m\gamma }\right) +\frac{\kappa }{m\gamma }$.

The form of interference manifests itself in $\rho \left( x,x,t\right) $ : 
\begin{equation}
\rho \left( x,x,t\right) =1+\exp \left( f_{1}\left( t\right) \left(
k_{1}-k_{2}\right) ^{2}\right) \left\{ \cos \left( \left[ k_{1}-k_{2}\right]
x-\frac{\left( k_{1}^{2}-k_{2}^{2}\right) }{2m\gamma }\left( 1-e^{-\gamma
t}\right) \right) \right\}  \label{interference2}
\end{equation}

Clearly from the form of $\ f_{1}\left( t\right) $ (\ref{soln1}) the above
indicates that the reduction in contrast is not exponential with $t $; hence
the interference is qualitatively of a milder form than that due to
conventional baths which includes the approach due to quantum semi-groups.

\section{Flavour oscillation}

In order to understand the effect of decoherence on flavour oscillations we
can for simplicity consider two flavours and flavours created with a sharp
momentum. Consequently for any momentum ket $\left| k\right\rangle $ there
will be two possible quantum numbers corresponding to the two mass
eigenstates with mass eigenvalues $m_{1}$ and $m_{2}$. To reflect this the
density matrix will have a $2\times 2$ matrix structure 
\begin{equation}
\rho =\rho _{0}\mathbf{1}+\rho _{i}\sigma _{i}  \label{density matrix}
\end{equation}%
where $\sigma _{i},i=1\ldots 3$ are the Pauli matrices and $\mathbf{1}$ is
the identity matrix and a summation convention over repeated indices is
assumed. In terms of \ 
\begin{equation}
m_{12}=\frac{m_{2}-m_{1}}{m_{2}+m_{1}},\frac{1}{m}=\frac{1}{2}\left( \frac{1%
}{m_{1}}+\frac{1}{m_{2}}\right)
\end{equation}%
we can write the kinetic energy as $H_{0}=\frac{\widehat{p}^{2}}{2m}\left( 
\mathbf{1}+m_{12}\sigma _{3}\right) $. Because the decoherence is
gravitational it will be sensitive to masses; hence we can take it to have
in general a structure with $r$ no longer a scalar function. A random $%
2\times 2$ matrix structure $r=r_{0}\mathbf{1}+r_{i}\sigma _{i}$ would be
natural. The previous master equation generalises to 
\begin{equation}
i\frac{d}{dt}\left( \rho _{0}\mathbf{1}+\rho _{j}\sigma _{j}\right)
=[H_{0},\rho _{0}\mathbf{1}+\rho _{j}\sigma _{j}]-i\Omega \left[ \widehat{p}%
\left( r_{0}\mathbf{1}+r_{i}\sigma _{i}\right) ,\left[ \widehat{p}\left(
r_{0}\mathbf{1}+r_{j}\sigma _{j}\right) ,\rho _{0}\mathbf{1}+\rho _{k}\sigma
_{k}\right] \right]  \label{master6}
\end{equation}

We now introduce 
\begin{equation}
\widetilde{\rho }_{\mu }\left( \kappa ,\Delta \right) =tr\left( \rho _{\mu }%
\widehat{D}\right)  \label{transform3}
\end{equation}
for $\mu =0,1,2,3.$

The equations arising for $\rho _{\mu }$ are given in the appendix. In order
to draw qualitative conclusions it is sufficient to simplify by requiring
only $r_{0}$ and $r_{3\mathrm{\ }}$ as being the only non-zero random
variables. The analysis then simplifies and is given in Appendix A.

The flavour mixing unitary matrix is%
\begin{equation}
U=\left( 
\begin{array}{cc}
\cos \theta & \sin \theta \\ 
-\sin \theta & \cos \theta%
\end{array}%
\right)
\end{equation}%
and relates flavour to mass eigenstates through $\left| \alpha \right\rangle
=U_{\alpha j}\left| j\right\rangle $ ( where the Latin indices apply to mass
eigenstates and Greek indices to flavour eigenstates). The initial density
matrix will represent a particle in flavour eigenstate 1 with momentum $p$.
Its matrix elements are%
\begin{eqnarray}
\rho _{11}\left( t=0\right) &=&\cos ^{2}\theta \,\,\left| p\right\rangle
\left\langle p\right|  \label{initialmatrix} \\
\rho _{22}\left( t=0\right) &=&\sin ^{2}\theta \,\,\left| p\right\rangle
\left\langle p\right|  \notag \\
\rho _{12}\left( t=0\right) &=&\sin \theta \cos \theta \,\,\left|
p\right\rangle \left\langle p\right|  \notag \\
\rho _{21}\left( t=0\right) &=&\sin \theta \cos \theta \,\,\left|
p\right\rangle \left\langle p\right| .  \notag
\end{eqnarray}

The corresponding elements for $\widetilde{\rho }$ are 
\begin{eqnarray*}
\widetilde{\rho }_{0}\left( t=0\right) &=&\pi e^{i\Delta \left( p+\kappa
/2\right) }\delta \left( \kappa \right) \\
\widetilde{\rho }_{3}\left( t=0\right) &=&\pi \cos \left( 2\theta \right)
e^{i\Delta \left( p+\kappa /2\right) }\delta \left( \kappa \right) \\
\widetilde{\rho }_{1}\left( t=0\right) &=&\pi \sin \left( 2\theta \right)
e^{i\Delta \left( p+\kappa /2\right) }\delta \left( \kappa \right) \\
\widetilde{\rho }_{2}\left( t=0\right) &=&0
\end{eqnarray*}

The solution at a general time (c.f. Appendix A) is%
\begin{eqnarray}
\widetilde{\rho }_{0}\left( t\right) &=&\frac{\pi }{2}e^{i\Delta \left(
p+\kappa /2\right) }\delta \left( \kappa \right) \left\{ 
\begin{array}{c}
\left( 1+\cos \left( 2\theta \right) \right) \exp \left[ -t\left( \left(
r_{0}+r_{3}\right) ^{2}\kappa ^{2}\frac{\mathcal{I}(t)}{t}-i\frac{\left(
1+m_{12}\right) \kappa \left( 2p+\kappa \right) }{2m}\right) \right] + \\ 
\left( 1-\cos \left( 2\theta \right) \right) \exp \left[ -t\left( \left(
r_{0}-r_{3}\right) ^{2}\kappa ^{2}\frac{\mathcal{I}(t)}{t}-i\frac{\left(
1-m_{12}\right) \kappa \left( 2p+\kappa \right) }{2m}\right) \right]%
\end{array}%
\right\}  \notag \\
\widetilde{\rho }_{3}\left( t\right) &=&\frac{\pi }{2}e^{i\Delta \left(
p+\kappa /2\right) }\delta \left( \kappa \right) \left\{ 
\begin{array}{c}
\left( 1+\cos \left( 2\theta \right) \right) \exp \left[ -t\left( \left(
r_{0}+r_{3}\right) ^{2}\kappa ^{2}\frac{\mathcal{I}(t)}{t}-i\frac{\left(
1+m_{12}\right) \kappa \left( 2p+\kappa \right) }{2m}\right) \right] - \\ 
\left( 1-\cos \left( 2\theta \right) \right) \exp \left[ -t\left( \left(
r_{0}-r_{3}\right) ^{2}\kappa ^{2}\frac{\mathcal{I}(t)}{t}-i\frac{\left(
1-m_{12}\right) \kappa \left( 2p+\kappa \right) }{2m}\right) \right]%
\end{array}%
\right\}  \notag \\
\widetilde{\rho }_{1}\left( t\right) &=&\pi \sin \left( 2\theta \right)
e^{i\Delta \left( p+\kappa /2\right) }\delta \left( \kappa \right) \cos
\left( \frac{m_{12}}{m}p^{2}t\right) \exp \left( -4r_{3}^{2}p^{2}\mathcal{I}%
(t)\right)  \notag \\
\widetilde{\rho }_{2}\left( t\right) &=&\pi \sin \left( 2\theta \right)
e^{i\Delta \left( p+\kappa /2\right) }\delta \left( \kappa \right) \sin
\left( \frac{m_{12}}{m}p^{2}t\right) \exp \left( -4r_{3}^{2}p^{2}\mathcal{I}%
(t)\right)  \label{results2}
\end{eqnarray}%
where (c.f. (\ref{arctan})): 
\begin{equation}
\mathcal{I}(t)\equiv \int_{0}^{t}\Omega (t^{\prime })dt^{\prime }=\Omega
_{0}t+\widetilde{\gamma }\mathrm{ln}(1+t/a)+\frac{\widetilde{\Gamma }}{\sqrt{%
b}}\mathrm{tan}^{-1}(\sqrt{b}t)  \label{arctan2}
\end{equation}

The probability for the transition of flavour 1 to flavour 2 at time $t$ is $%
tr\left( \rho _{f\mu }\rho _{\mu }\left( t\right) \right) $ where 
\begin{equation*}
\rho _{f0}=\frac{1}{2}\left| p\right\rangle \left\langle p\right| ,\;\rho
_{f1}=-\frac{1}{2}\sin 2\theta \;\left| p\right\rangle \left\langle p\right|
,\;\rho _{f2}=0,\;\rho _{f3}=-\frac{1}{2}\cos 2\theta \mathrm{{\ }\left|
p\right\rangle \left\langle p\right| . }
\end{equation*}

{}From (\ref{results2}),(\ref{arctan2}) it is clear that decoherence affects
this probability with an exponential damping \emph{only if} the cosmological
term $\Omega $ is the \emph{constant }$\Omega _{0}$. In particular, in the
absence of the $\Omega _{0}$ term, and in the limit $\widetilde{\Gamma }=0$,
the decoherence due to the D-particle foam results in power damping for
large times $t\rightarrow \infty $, the terms $\widetilde{\rho }_{i}\sim
t^{-\delta _{i}^{2}}$, $i=0,3$, while $\widetilde{\rho }_{i}\sim
t^{-p^{2}\delta _{i}^{2}}$, $i=1,2$, i.e. the scaling power depends on the
probe's momentum (with $\delta _{i}$ appropriate constants depending on the
term we look at in (\ref{arctan2}). With the general $\Omega \left( t\right) $
of (\ref{cosmological}) the probability for flavour oscillation, $%
P_{1\rightarrow 2}$, is proportional to 
\begin{equation}
P_{1\rightarrow 2}\propto \left( \sin 2\theta \right) ^{2}\left( 1-\exp 
\left[ -4\sigma ^{2}p^{2}\mathcal{I}(t)\right] \right) \cos \left( \frac{%
m_{12}p^{2}t}{m}\right) ~.  \label{flavourprob}
\end{equation}%
with ${}(t)$ given by (\ref{arctan2}).

We remark at this stage that, 
in view of the fundamental CPT violation inherent in the 
master equation (\ref{master3}) or (\ref{master5}), 
as a result of the microscopic time irreversibility under $t \to -t $
of the entanglement terms of this equation, it is also
possible (but not necessary) to consider models of D-foam 
in which the momentum
transfer (\ref{trnsf}), (\ref{random}) 
differs between particle and antiparticle sectors.
In such a case, one would then obtain different decoherence
coefficients for particles and antiparticles, a model
which was already used in \cite{barenboim} in order to 
fit LSND data~\cite{reviewnu}.

Although our simplified model for flavour oscillations, using a
non-relativistic bosonic two-level system, is too simplified for realistic
phenomenology of (neutrino) flavour oscillations, nevertheless we believe
that it captures the basic features induced by Liouville-decoherence, which
are expected to persist in the relativistic neutrino case. The important
point is the damping factor $\exp \left[ -4\sigma ^{2}p^{2}\mathcal{I}(t)%
\right] $ in front of the oscillatory term, which depends on the square of
the momentum of the probe, as well as the space-time foam characteristics,
such as the dispersion $\sigma ^2$ (\ref{random}) in the momentum transfer
during the interaction of the matter probe with the D-particle defect.

In this respect, it is interesting to compare the general form (\ref%
{flavourprob}) of the Liouville decoherence, in which the entanglement is
time dependent in general, with the generic models of Lindblad decoherence,
based on the approach of \cite{ehns}, in which the environmental
entanglement is characterized by a constant decoherence matrix. For
instance, consider for definiteness the two-flavour oscillation case, in the
completely-positive neutrino decoherence model of \cite{lisi,benatti}, for
which there is only one real and positive, constant in time, but possibly
probe-energy dependent, decoherence parameter $\gamma $ to characterize the
quantum-gravity entanglement. In the case where the energy and lepton number
of the neutrino are assumed conserved (on average, at least) in the presence
of quantum-gravity fluctuations, the oscillation probability reads in that
model: 
\begin{equation}
P_{\nu _{\alpha }\rightarrow \nu _{\beta }}=\frac{1}{2}\left( \sin 2\theta
\right) ^{2}\left( 1-\exp (-\gamma L)\mathrm{cos}(\frac{%
(m_{1}^{2}-m_{2}^{2})L}{2E_{\nu }})\right)  \label{neutrino}
\end{equation}%
where $L=t$ is the neutrino oscillation length, and $E_{\nu }$ is the
average energy of the neutrino beam.

First of all let us concentrate in the form of the oscillatory terms. To
understand the difference between our case (\ref{flavourprob}) and 
(\ref{neutrino}), we should remind the reader that flavour oscillations can be
analyzed in two ways: the first involves flavours with sharp momentum, in
which case one considers oscillations in time, as done in the present work,
whilst the other involves flavour with sharp energy, in which case the
oscillations are in space. This is the conventional neutrino case, leading
to (\ref{neutrino}). The conclusions should not change, however,
qualitatively. For our purposes it is not necessary, therefore, to consider
the second case in order to get a qualitative description of the D-foam
effects. This will be important only when we embark on a detailed
phenomenology, by looking at specific experiments, in which case
complications involving wave packets and the joint presence of the above
effects must be incorporated. One should also take into account that the
neutrino is a highly relativistic system, with a very small mass, in
contrast to our non-relativistic bosonic system examined in this section for
simplicity. One expects, therefore, that when we consider in our approach
relativistic systems, with sharp energy, a similar form for the oscillatory
term as in (\ref{neutrino}) will be obtained. Qualitatively, however, in
both (\ref{flavourprob}) and (\ref{neutrino}), the main reason for
oscillations is the conventional mass difference between the mass
eigenstates, which agrees with the present phenomenology.

The exponential damping (with time or oscillation length) induced by the
decoherence term is only present in our case in the case of a cosmological
constant. The generic D-foam effects are time dependent in the way indicated
in (\ref{arctan2}), which does not introduce exponential damping. The
important feature, however, is the quadratic momentum dependence $p^2$ of
the exponent of these ``damping'' terms, which persists in the relativistic
case, since it originates from the specific $[p, [p, \rho]]$ diffusion term
of the master equation (\ref{master5}), as well as their dependence on the
specific characteristics of the foam, such as the dispersion $\sigma$ (\ref%
{random}) for the momentum transfer during the interaction of the particle
probe with the foam. Experimentally, therefore, one should be able to bound
this dispersion by comparing the oscillation probability (\ref{flavourprob})
as a function of the oscillation length (time) with the corresponding
experimental curve. Currently there are stringent limits from neutrino
physics on quantum decoherence terms~\cite{winstan}, and one can envisage
that such bounds provide important restrictions on our D-particle foam
characteristic dispersion $\sigma$, when the model is applied to
relativistic neutrino beams. We postpone a detailed phenomenological
analysis of our effects for a future publication, due to the complications
mentioned above.

It is also worth noticing at this stage that damping factors, which resemble
those due to decoherence, can arise simply~\cite{olsson} by conventional
uncertainties in the energy (or momentum, depending of the formalism
adopted) of the neutrino beam. If $\sigma _{E}$ denotes the corresponding
dispersion, then, for the two-flavour oscillation problem one finds: 
\begin{equation}
P_{\nu _{\alpha }\rightarrow \nu _{\beta }}=\frac{1}{2}\left( \sin 2\theta
\right) ^{2}\left( 1-\exp (-2\sigma _{E}^{2}(m_{1}^{2}-m_{2}^{2})^{2})%
\mathrm{cos}(\frac{(m_{1}^{2}-m_{2}^{2})\langle L\rangle }{2\langle E_{\nu
}\rangle })\right)  \label{gaussian}
\end{equation}%
The reader is invited to compare (\ref{gaussian}) with (\ref{flavourprob})
and (\ref{neutrino}). We note that by writing~\cite{olsson}: $\sigma _{E}=%
\frac{L}{4E_{\nu }}r$, with $r\sim \delta E/E$ (ignoring, as negligible,
uncertainties in oscillation lengths), then one may rewrite (\ref{gaussian})
in a decoherence form (\ref{neutrino}) with the decoherence coefficient $%
\gamma \sim (m_{1}^{2}-m_{2}^{2})^{2}Lr^{2}/8E_{\nu }^{2}$. For atmospheric
neutrinos, this yields the bound $\gamma _{\mathrm{atm}}\leq 10^{-24}$~GeV,
assuming $r^{2}={}(1)$. Similar bounds can be obtained for the D-foam
decoherence damping factor (\ref{flavourprob}), which notably is independent
of the mass difference. Indeed, one can also formally rewrite the damping
factor in (\ref{flavourprob}) as a decoherence (\ref{neutrino}) term, with $%
\gamma =4\sigma ^{2}p^{2}{}I(L)/L$, with $L=t$. However, for reasons
explained above this formula is not quite precise, and for quantitative
comparison one should repeat the analysis with relativistic neutrino
systems, in the sharp energy formalism. This is left for future work. 

Finally, before closing this section we would like to compare
our result (\ref{flavourprob}) with the energy-driven Lindblad decoherence
models for two-flavour oscillations considered in \cite{adler}. 
In the latter case, the pertinent master equation for the density matrix
of matter propagating in a self-adjoint Lindblad environment, spanned
by operators $D_n^\dagger = D_n$, with $[D_n , H] = 0$, 
reads (in units $\hbar =1=c$):
\begin{equation}
\partial_t \rho = i[\rho, H] - \sum_n [ D_n, [D_n, \rho]]~.
\label{adler1}
\end{equation}
When specialized to a two-level system, Adler~\cite{adler} asserts
that the only possible choices that commute with $H$ 
are $D_n=\kappa_n \widehat 1$,
with $\widehat 1$ the identity operator, 
and $D_n = \lambda_n H$, with $\kappa_n, \lambda_n$ appropriate constants.
In the relativistic massive neutrino case, 
with fixed energy $E$ for the neutrino beam,
$H \simeq p + \frac{m^2}{2p}$, from which, 
by projecting onto mass eigenstates, and  
concentrating on the non-diagonal density-matrix elements, 
Eq.~(\ref{adler1}) 
yields 
the following leading-order estimate for the decoherence coefficient
$\gamma$~\cite{adler}:
\begin{equation}\label{adler2}
\gamma \sim \frac{(m_1^2 - m_2^2)^2}{4E^2 M_P}
\end{equation}
where $1/M_P = \sum_n \lambda_n^2 $ is a characteristic Quantum Gravity 
scale.
In the non-relativistic case at hand, from the rest-mass terms in the 
expansion of the Hamiltonian $H \simeq m + \frac{p^2}{2m}$, 
we obtain the following estimate for the decoherence term
$\gamma_{nr} \sim (m_1 - m_2)^2/M_P + {\cal O}(p^4 (m_1 - m_2)^2/4m_1^2m_2^2M_P)$.
These estimates are much more suppressed.than the estimates of $\cite{lisi}$,
who allowed for the decoherence coefficients (in the relativistic case) 
to be of order 
$E^2/M_P$, with $E$ a fixed neutrino energy. 

We now note that this latter, quadratic in energy, dependence 
of the decoherence coefficient 
appears to be the case of our Liouville decoherence,
and owes its form to the peculiar (quadratic) momentum dependence of
the Liouville environmental decoherening entanglement term of the 
master equation (\ref{master5}). The form of this term 
is independent of the relativistic or non-relativistic nature of the
Hamiltonian $H$, whilst the form of (\ref{adler1}),(\ref{adler2}) 
depends crucially on the form of the Hamiltonian $H$. 

In the Liouville decoherence case of the D-particle foam, examined 
in the present work, the Lindblads, 
given by $\overline u \propto r \widehat p$ (c.f. (\ref{trnsf})), 
although self-adjoint and commuting
with the Hamiltonian of the non-relativistic system, 
nevertheless are \emph{not} proportional to it. 
This is due to the fact that the Lindblad operators in this case
are proportional to the metric tensor (\ref{master3}),  as a result of the 
underlying general coordinate invariance of the Liouville-string system, 
which is not 
taken into account in the simple quantum mechanical case of 
\cite{adler}. 
This leads to an entirely different energy (momentum)
dependence ($\propto p^2$) of the decoherence coefficient in the 
Liouville case, as we have seen above. 
With
these comments we conclude our phenomenological analysis on Liouville
decoherence.

\section{Conclusions}

In this work we have examined phenomenological consequences of a decoherence
master equation describing the propagation of low-energy string matter in a
space-time foam background. The basic machinery was the Liouville-string
approach to decoherence.

We have restricted ourselves in a specific model for the space time foam,
involving the interaction of strings with space-time D-particle defects. We
have seen that the pertinent master equation exhibits diffusion terms and
decoherence, but that there is not always exponential damping. The reason is
the time dependence of some of the decoherence terms. The Liouville
decoherence involves a term which is quadratic in the momentum transfer
encountered during the interaction of the particle probe with the space-time
defect. By considering a specific model of random foam, characterized by the
dispersion of such a momentum transfer, we have been able to solve the
pertinent decoherence equation rather simply, and give the form of the
corresponding probability for the oscillation from one flavour to another,
in a toy non-relativistic bosonic model of two flavour oscillations. Our
formalism also allows for the consideration of possible decoherence effects
arising from dark energy contributions in the space time. Such effects have
also been taken into account in the expression for the final oscillation
probability

Although the model is not strictly appropriate for neutrino physics,
nevertheless it exhibits the basic properties of Liouville decoherence that
one expects to persist in the fully relativistic neutrino case. The basic
feature is a damping looking factor $\exp (-4p^{2}\sigma ^{2}{\mathcal I}(t))$, in
front of the conventional oscillation terms that are due to the mass
difference of the pertinent mass eigenstates. The damping factor depends on
the characteristics of the foam, but leads to exponential damping (with the
time $t$) only in case there is a cosmological \emph{constant} term in space
time.

The simplified analysis presented in the current article demonstrates that
generic phenomenological analyses, based on constant decoherence
coefficients, might be misleading, yielding sometimes incorrect bounds of
the relevant space-time foam effects. One should resort, whenever possible,
to rather detailed microscopic models of space-time foam, as we have done
here, before embarking on detailed phenomenological analyses. One should
also be careful to disentangle conventional matter effects, or effects
related simply to uncertainties in the energy or oscillation lengths, before
concluding on the possible r\^ole of fundamental physics effects in particle
processes, such as flavour oscillation. We hope to return to a detailed
phenomenological analysis of D-particle foam effects in neutrinos and other
particle probes, taking into account of the appropriate spin structures as
well, in a future publication.

\section*{Acknowledgements}

We acknowledge discussions with G. Barenboim and A. Waldron-Lauda.

\section*{Appendix A: Detailed solution of the master equation}

By multiplying (\ref{master6}) with $\sigma _{\mu }$ and taking traces the
following equations arise:%
\begin{equation}
\frac{\partial }{\partial t}\widetilde{\rho }_{0}=\frac{\kappa }{m}\left( 
\frac{\partial }{\partial \Delta }\widetilde{\rho }_{0}+m_{12}\frac{\partial 
}{\partial \Delta }\widetilde{\rho }_{3}\right) -\Omega \kappa ^{2}\left(
r_{\mu }r_{\mu }\widetilde{\rho }_{0}+2r_{0}r_{j}\widetilde{\rho }%
_{j}\right) \mathrm{\ \ \ \ \ \ \ }  \label{A1}
\end{equation}

and%
\begin{eqnarray}
\frac{\partial }{\partial t}\widetilde{\rho }_{j} &=&\left( \frac{m_{12}}{m}%
\varepsilon _{3lj}+4\Omega h_{lj}\right) \left( i\frac{\partial }{\partial
\Delta }+\frac{\kappa }{2}\right) ^{2}\widetilde{\rho }_{l}-4i\Omega \kappa
f_{lj}\left( i\frac{\partial }{\partial \Delta }+\frac{\kappa }{2}\right) 
\widetilde{\rho }_{l}+  \notag \\
&&\frac{\kappa }{m}\left( m_{12}\delta _{j3}\frac{\partial }{\partial \Delta 
}\widetilde{\rho }_{0}+\left( \delta _{ij}-im_{12}\varepsilon _{3lj}\right) 
\frac{\partial }{\partial \Delta }\widetilde{\rho }_{l}\right) -  \notag \\
&&\Omega \kappa ^{2}\left( 2r_{0}r_{j}\widetilde{\rho }_{0}+\left(
r_{l}r_{j}+r_{0}^{2}\delta _{jl}+ir_{0}\varepsilon
_{lkj}r_{k}-if_{lj}\right) \widetilde{\rho }_{l}\right)  \label{A2}
\end{eqnarray}
where $f_{jp}=r_{0}r_{s}\varepsilon _{jps}+ir_{l}r_{l}\delta
_{jp}-ir_{j}r_{p}$.

When $r_{0}$ and $r_{3}$ are the only non-zero components of $r_{\mu }$ then
the above equations become 
\begin{eqnarray}
\frac{\partial }{\partial t}\widetilde{\rho }_{0} &=&\frac{\kappa }{m}\left( 
\frac{\partial }{\partial \Delta }\widetilde{\rho }_{0}+m_{12}\frac{\partial 
}{\partial \Delta }\widetilde{\rho }_{3}\right) -\Omega \kappa ^{2}\left(
\left( r_{0}^{2}+r_{3}^{2}\right) \widetilde{\rho }_{0}+2r_{0}r_{3}%
\widetilde{\rho }_{3}\right)  \label{A3} \\
\frac{\partial }{\partial t}\widetilde{\rho }_{1} &=&-\left( i\frac{\partial 
}{\partial \Delta }+\frac{\kappa }{2}\right) ^{2}\left( \frac{m_{12}}{m}%
\widetilde{\rho }_{2}+4\Omega r_{3}^{2}\widetilde{\rho }_{1}\right)
-4i\Omega \kappa \left( i\frac{\partial }{\partial \Delta }+\frac{\kappa }{2}%
\right) \left( -r_{0}r_{3}\widetilde{\rho }_{2}+ir_{3}^{2}\widetilde{\rho }%
_{1}\right) +  \notag \\
&&\frac{\kappa }{m}\left( \frac{\partial }{\partial \Delta }\widetilde{\rho }%
_{1}+im_{12}\frac{\partial }{\partial \Delta }\widetilde{\rho }_{2}\right)
-\Omega \kappa ^{2}\left( \left( r_{0}^{2}+r_{3}^{2}\right) \widetilde{\rho }%
_{1}+2ir_{0}r_{3}\widetilde{\rho }_{2}\right)  \label{A4} \\
\frac{\partial }{\partial t}\widetilde{\rho }_{2} &=&\left( i\frac{\partial 
}{\partial \Delta }+\frac{\kappa }{2}\right) ^{2}\left( \frac{m_{12}}{m}%
\widetilde{\rho }_{1}-4\Omega r_{3}^{2}\widetilde{\rho }_{2}\right)
-4i\Omega \kappa \left( i\frac{\partial }{\partial \Delta }+\frac{\kappa }{2}%
\right) \left( r_{0}r_{3}\widetilde{\rho }_{1}+ir_{3}^{2}\widetilde{\rho }%
_{2}\right) +  \label{A6} \\
&&\frac{\kappa }{m}\left( \frac{\partial }{\partial \Delta }\widetilde{\rho }%
_{2}-im_{12}\frac{\partial }{\partial \Delta }\widetilde{\rho }_{1}\right)
-\Omega \kappa ^{2}\left( \left( r_{0}^{2}+r_{3}^{2}\right) \widetilde{\rho }%
_{2}-2ir_{0}r_{3}\widetilde{\rho }_{1}\right)  \notag \\
\frac{\partial }{\partial t}\widetilde{\rho }_{3} &=&\frac{\kappa }{m}\left( 
\frac{\partial }{\partial \Delta }\widetilde{\rho }_{3}+m_{12}\frac{\partial 
}{\partial \Delta }\widetilde{\rho }_{0}\right) -\Omega \kappa ^{2}\left(
2r_{0}r_{3}\widetilde{\rho }_{0}+\left( r_{0}^{2}+r_{3}^{2}\right) 
\widetilde{\rho }_{3}\right)  \label{A7}
\end{eqnarray}

The equations (\ref{A3}), (\ref{A4}), (\ref{A6}) and (\ref{A7}) form two
decoupled sets, one involving $\widetilde{\rho }_{0}$ and $\widetilde{\rho }%
_{3}$, the other $\widetilde{\rho }_{1}$ and $\widetilde{\rho }_{2}$. In
order to solve these equations we introduce the Fourier transforms $%
\widetilde{\widetilde{\rho }}_{\mu }$ defined by 
\begin{equation}
\widetilde{\widetilde{\rho }}_{\mu }\left( \kappa ,\eta \right)
=\int_{-\infty }^{\infty }\widetilde{\rho }_{\mu }\left( \kappa ,\Delta
\right) e^{-2\pi i\eta \Delta }d\Delta
\end{equation}%
(so that $\widetilde{\rho }_{\mu }\left( \kappa ,\Delta \right)
=\int_{-\infty }^{\infty }\widetilde{\widetilde{\rho }}_{\mu }\left( \kappa
,\eta \right) e^{2\pi i\eta \Delta }d\eta )$.

We then have 
\begin{equation}
i\frac{\partial }{\partial t}\left( 
\begin{array}{c}
\widetilde{\widetilde{\rho }}_{0} \\ 
\widetilde{\widetilde{\rho }}_{3}%
\end{array}%
\right) =M\left( 
\begin{array}{c}
\widetilde{\widetilde{\rho }}_{0} \\ 
\widetilde{\widetilde{\rho }}_{3}%
\end{array}%
\right)
\end{equation}%
where 
\begin{equation*}
M=\left( 
\begin{array}{cc}
-\frac{2\pi \kappa \eta }{m}-i\Omega \kappa ^{2}\left(
r_{o}^{2}+r_{3}^{2}\right) & -\frac{2\pi \kappa \eta m_{12}}{m}-2i\Omega
\kappa ^{2}r_{0}r_{3} \\ 
-\frac{2\pi \kappa \eta m_{12}}{m}-2i\Omega \kappa ^{2}r_{0}r_{3} & -\frac{%
2\pi \kappa \eta }{m}-i\Omega \kappa ^{2}\left( r_{o}^{2}+r_{3}^{2}\right)%
\end{array}%
\right) .
\end{equation*}%
Now for $\mathcal{C}=\left( 
\begin{array}{cc}
1 & -1 \\ 
1 & 1%
\end{array}%
\right) $%
\begin{equation*}
\mathcal{C}^{-1}M\mathcal{C=}\left( 
\begin{array}{cc}
\lambda _{1} & 0 \\ 
0 & \lambda _{2}%
\end{array}%
\right)
\end{equation*}%
where 
\begin{eqnarray}
\lambda _{1} &=&-\frac{\kappa }{m}\left( 2\left( 1+m_{12}\right) \pi \eta
+im\left( r_{0}+r_{3}\right) ^{2}\kappa \Omega \right) \\
\lambda _{2} &=&-\frac{\kappa }{m}\left( 2\left( 1-m_{12}\right) \pi \eta
+im\left( r_{0}-r_{3}\right) ^{2}\kappa \Omega \right) .  \notag
\end{eqnarray}
Notice that so far $\Omega$ has been kept arbitrary, in other words it may
be in general a function of (cosmic) time $\Omega (t)$.

On introducing 
\begin{equation*}
\left( 
\begin{array}{c}
\widehat{\rho }_{0} \\ 
\widehat{\rho }_{3}%
\end{array}%
\right) =\mathcal{C}^{-1}\left( 
\begin{array}{c}
\widetilde{\widetilde{\rho }}_{0} \\ 
\widetilde{\widetilde{\rho }}_{3}%
\end{array}%
\right)
\end{equation*}%
we have 
\begin{eqnarray}
\widehat{\rho }_{0}\left( t\right) &=&e^{-i\int_{0}^{t}\lambda
_{1}(t^{\prime })dt^{\prime }}\widehat{\rho }_{0}\left( 0\right) \\
\widehat{\rho }_{3}\left( t\right) &=&e^{-i\int_{0}^{t}\lambda
_{2}(t^{\prime })dt^{\prime }}\widehat{\rho }_{3}\left( 0\right) .  \notag
\end{eqnarray}%
Similarly 
\begin{equation*}
i\frac{\partial }{\partial t}\left( 
\begin{array}{c}
\widetilde{\widetilde{\rho }}_{1} \\ 
\widetilde{\widetilde{\rho }}_{2}%
\end{array}%
\right) =N\left( 
\begin{array}{c}
\widetilde{\widetilde{\rho }}_{1} \\ 
\widetilde{\widetilde{\rho }}_{2}%
\end{array}%
\right)
\end{equation*}%
where 
\begin{equation*}
N_{11}=N_{22}=-\frac{2\pi \eta \kappa }{m}-16i\pi ^{2}r_{3}^{2}\eta
^{2}\Omega +ir_{3}^{2}\kappa ^{2}\Omega -i\left( r_{0}^{2}+r_{3}^{2}\right)
\kappa ^{2}\Omega
\end{equation*}%
and 
\begin{equation*}
N_{21}=-N_{12}=\frac{4im_{12}\pi ^{2}\eta ^{2}}{m}+\frac{im_{12}\kappa ^{2}}{%
4m}-8\pi r_{0}r_{3}\eta \kappa \Omega .
\end{equation*}%
For $\mathcal{D}=\left( 
\begin{array}{cc}
-i & i \\ 
1 & 1%
\end{array}%
\right) $%
\begin{equation*}
\mathcal{D}^{-1}N\mathcal{D=}\left( 
\begin{array}{cc}
\mu _{1} & 0 \\ 
0 & \mu _{2}%
\end{array}%
\right)
\end{equation*}%
and so, on defining 
\begin{equation*}
\left( 
\begin{array}{c}
\widehat{\rho }_{1} \\ 
\widehat{\rho }_{2}%
\end{array}%
\right) =\mathcal{D}^{-1}\left( 
\begin{array}{c}
\widetilde{\widetilde{\rho }}_{1} \\ 
\widetilde{\widetilde{\rho }}_{2}%
\end{array}%
\right)
\end{equation*}%
we have%
\begin{eqnarray}
\widehat{\rho }_{1}\left( t\right) &=&e^{-i\int_{0}^{t}\mu _{1}(t^{\prime
})dt^{\prime }}\widehat{\rho }_{1}\left( 0\right) \\
\widehat{\rho }_{2}\left( t\right) &=&e^{-i\int_{0}^{t}\mu _{2}(t^{\prime
})dt^{\prime }}\widehat{\rho }_{2}\left( 0\right) .  \notag
\end{eqnarray}%
This analysis can be summarised as follows:%
\begin{eqnarray}
\widetilde{\widetilde{\rho }}_{0}\left( t\right) &=&\frac{1}{2}\left(
e^{-i\int_{0}^{t}\lambda _{1}(t^{\prime })dt^{\prime
}}+e^{-i\int_{0}^{t}\lambda _{2}(t^{\prime })dt^{\prime }}\right) \widetilde{%
\widetilde{\rho }}_{0}\left( 0\right) +\frac{1}{2}\left(
e^{-i\int_{0}^{t}\lambda _{1}(t^{\prime })dt^{\prime
}}-e^{-i\int_{0}^{t}\lambda _{2}(t^{\prime })dt^{\prime }}\right) \widetilde{%
\widetilde{\rho }}_{3}\left( 0\right)  \notag \\
\widetilde{\widetilde{\rho }}_{3}\left( t\right) &=&\frac{1}{2}\left(
e^{-i\int_{0}^{t}\lambda _{1}(t^{\prime })dt^{\prime
}}-e^{-i\int_{0}^{t}\lambda _{2}(t^{\prime })dt^{\prime }}\right) \widetilde{%
\widetilde{\rho }}_{0}\left( 0\right) +\frac{1}{2}\left(
e^{-i\int_{0}^{t}\lambda _{1}(t^{\prime })dt^{\prime
}}+e^{-i\int_{0}^{t}\lambda _{2}(t^{\prime })dt^{\prime }}\right) \widetilde{%
\widetilde{\rho }}_{3}\left( 0\right)  \label{results} \\
\widetilde{\widetilde{\rho }}_{1}\left( t\right) &=&\frac{i}{2}\left(
-i\left( e^{-i\int_{0}^{t}\mu _{1}(t^{\prime })dt^{\prime
}}+e^{-i\int_{0}^{t}\mu _{2}(t^{\prime })dt^{\prime }}\right) \widetilde{%
\widetilde{\rho }}_{1}\left( 0\right) +\left( e^{-i\int_{0}^{t}\mu
_{2}(t^{\prime })dt^{\prime }}-e^{-i\int_{0}^{t}\mu _{1}(t^{\prime
})dt^{\prime }}\right) \widetilde{\widetilde{\rho }}_{2}\left( 0\right)
\right)  \notag \\
\widetilde{\widetilde{\rho }}_{2}\left( t\right) &=&\frac{i}{2}\left(
-\left( e^{-i\int_{0}^{t}\mu _{2}(t^{\prime })dt^{\prime
}}-e^{-i\int_{0}^{t}\mu _{1}(t^{\prime })dt^{\prime }}\right) \widetilde{%
\widetilde{\rho }}_{1}\left( 0\right) -i\left( e^{-i\int_{0}^{t}\mu
_{1}(t^{\prime })dt^{\prime }}+e^{-i\int_{0}^{t}\mu _{2}(t^{\prime
})dt^{\prime }}\right) \widetilde{\widetilde{\rho }}_{2}\left( 0\right)
\right)  \notag
\end{eqnarray}%
For the specific case where $\Omega (t)$ has the time dependence indicated
in (\ref{cosmological}), the corresponding time integral is elementary: 
\begin{equation}
\mathcal{I}(t)\equiv \int_{0}^{t}\Omega (t^{\prime })dt^{\prime }=\Omega
_{0}t+\widetilde{\gamma }\mathrm{ln}(1+t/a)+\frac{\widetilde{\Gamma }}{\sqrt{%
b}}\mathrm{tan}^{-1}(\sqrt{b}t)  \label{arctan}
\end{equation}%
which, in the limit of constant $\Omega $, i.e. $\widetilde{\Gamma },%
\widetilde{\gamma }\rightarrow 0$ yields the standard term $\mathcal{I}(t,%
\widetilde{\gamma }=0,\widetilde{\Gamma }=0)=\Omega _{0}t$, responsible for
the usual exponential damping. The above results are used in section 4,
where we discuss Liouville decoherence in a toy two-generation oscillation
system.

\section*{Appendix B: Calculation of back reaction in D-particle foam}

\subsection*{D-particle foam contributions to master equation for
Liouville-decoherence}

The material in this Appendix is a review based on \cite{kmw}, where we
refer the reader for further details. Let us consider a $D$-particle,
located at $y^{i}(t=0)\equiv y_{i}$ of the spatial coordinates of a $(d+1)$%
-dimensional space time (which could be a D3-brane world), which at a time $%
t=0$ experiences an impulse, as a result of scattering with a matter string
state (see Fig. \ref{fig:dfoam}). In a $\sigma $-model framework, the
trajectory of the $D$-particle $y^{i}(t)$, $i=1,2,\dots d,$ a spatial index,
is described by inserting the following vertex operator in the $\sigma $%
-model of a free string: 
\begin{equation}
V=\int_{\partial \Sigma }g_{ij}y^{j}(t)\partial _{n}X^{i}  \label{path1}
\end{equation}%
where $g_{ij}$ denotes the spatial components of the metric, $\partial
\Sigma $ denotes the world-sheet boundary, $\partial _{n}$ is a normal
world-sheet derivative, $X^{i}$ are $\sigma $-model fields obeying Dirichlet
boundary conditions on the world sheet, and $t$ is a $\sigma $-model field
obeying Neumann boundary conditions on the world sheet, whose zero mode is
the target time. The space-time prior to Liouville dressing is assumed
Euclidean for formal reasons (convergence of the corresponding $\sigma $%
-model path integral). We note, however, that the final Liouville-dressed
target space-time acquires Minkowski signature as a result of the time-like
signature of the Liouville mode~\cite{emn,gravanis}.

In the non-relativistic approximation, appropriate for a heavy D-particle
defect of mass $M_s/g_s$, with $M_s$ the string scale, and $g_s$ the string
coupling, assumed weak ($g_s \ll 1$), the path $y^i(t)$ corresponding to the
impulse is given by: 
\begin{eqnarray}  \label{path}
y_i(t) &=& \left( \varepsilon y_i + u_i t \right)\Theta_\varepsilon (t) 
\notag \\
u_{i} &=&\left( k_{1}-k_{2}\right) _{i}~,
\end{eqnarray}
with $k_{1}\left( k_{2}\right) $ the momentum of the propagating string
state before (after) the recoil (see fig. \ref{fig:dfoam}); $y_{i}$ are the
spatial collective coordinates of the D particle, and the regularized
Heaviside functional operator $\Theta_\varepsilon (t)$ is given by (\ref%
{heaviside}) in the text~\cite{kmw}: 
\begin{eqnarray}  \label{heaviside2}
\Theta _{\varepsilon }\left( t\right) &=&\frac{1}{2\pi i} \int_{-\infty}^{%
\infty }\frac{dq}{q-i\varepsilon }e^{iqt},
\end{eqnarray}
Eq. (\ref{path}) contains actually a \emph{pair} of deformations
corresponding to the $\sigma$-model couplings $y_i$ and $u_i$. These
deformations are relevant in a world-sheet renormalization-group sense,
having anomalous scaling dimension $-\frac{\varepsilon^2}{2}$, i.e. to
leading order in a coupling constant expansion their renormalization-group $%
\beta$-functions read: 
\begin{equation}
\beta^{y^i} = -\frac{\varepsilon^2}{2}y^i~,\qquad \beta^{u^i} = -\frac{%
\varepsilon^2}{2}u^i~.  \label{betafnct}
\end{equation}
The deformations form a logarithmic conformal algebra (superconformal
algebra in the case of superstrings) which \emph{closes} if and only if one
identifies~\cite{kmw} the regulating parameter $\varepsilon^{-2}$ with the
world-sheet renormalization-group scale $\mathrm{ln}|L/a|^2$ ($L(a)$ is the
Infrared (Ultraviolet) world-sheet scale): 
\begin{equation}
\varepsilon^{-2} = \eta \mathrm{ln}|L/a|^2  \label{elog}
\end{equation}
where $\eta$ denotes the signature of time $t$ of the target-space manifold
of the $\sigma$-model (prior to Liouville dressing). For Euclidean
manifolds, assumed here for path-integral convergence, $\eta = +1$.

Upon the identification (\ref{elog}) the re-scaled couplings $\overline{y}_i
\equiv \frac{y_i}{\varepsilon}$ and $\overline{u}_i \equiv \frac{u_i}{%
\varepsilon}$ are \emph{marginal}, that is independent of the scale $%
\varepsilon$. It is these marginal couplings that are connected to
target-space quantities of physical significance, such as the space-time
back reaction of recoil, which we now proceed to calculate.

In what follows we shall concentrate only on the limit $\varepsilon
\rightarrow 0$. In this limit the dominant contributions come from the $%
u_{i} $ recoil deformation in (\ref{path}), which we shall restrict our
attention to from now on. The corresponding two-point correlation function
of the vertex operator associated with the deformation $u_{i}$
(Zamolodchikov metric in $u$-space $\mathcal{G}_{uu}$) has the leading-order
behaviour~\cite{kmw}: 
\begin{equation}
\mathcal{G}_{uu}\sim \frac{1}{\varepsilon ^{2}}+\mathrm{finite~terms~as~}%
\varepsilon \rightarrow 0  \label{guu}
\end{equation}%
The corresponding deformations contribute the following terms in the master
equation (\ref{master}): 
\begin{eqnarray}
\mathrm{Liouville~Entanglement~in~Eq.~(\ref{master})} &=&:\beta ^{u_{i}}%
\mathcal{G}_{u_{i}u_{j}}[u^{j},\rho ]:=-\frac{\varepsilon ^{2}}{2}[u_{i}%
\frac{1}{\varepsilon ^{2}}\delta _{ij},~[u^{j},\rho ]]=  \notag
\label{entangl} \\
&~&-\frac{1}{2}[u_{i},[u^{i},\rho ]]=-\varepsilon ^{2}\frac{1}{2}[\overline{u%
}_{i},[\overline{u}^{i},\rho ]]
\end{eqnarray}%
upon selecting antisymmetric operator ordering to ensure the validity of the
Lindblad properties of the evolution, and expressing the final result in
terms of marginal $\overline{u}$ velocities. The (relative) negative sign in
front of the entanglement term in the Liouville master equation (\ref%
{entangl}) is important for exponential damping, and it was assumed in the
text. Finally one identifies $\varepsilon ^{2}$ with the inverse of the
(Minkowski) target time $1/t$.

This procedure will therefore yield a master equation for Liouville
decoherence in a Minkowski space-time environment of the form: 
\begin{equation}\label{liouvdecohmaster1}
\overset{.}{\partial _{t}\rho _{Matter}}=i\left[ \rho _{Matter},H\right] -\,%
\frac{\widetilde{\gamma }}{t}\left[ \overline{u}_{j},\left[ \overline{u}%
^{j},\rho _{Matter}\right] \right] ~.
\end{equation}
where the (positive) constant $\widetilde{\gamma }$ has been incorporated in
order to take into account situations in which the density of D-particles on
the brane world is less than one per Planck volume, assumed above. Lacking a
detailed microscopic theory, which would in principle determine this density
from first principles, we have to resort to phenomenological considerations,
which are subject in principle to experimental tests. Thus, for us the
parameter $0<\widetilde{\gamma }<1$ will be a free parameter to be bounded
by experiment.

In the above derivation the identification $t\sim 1/\varepsilon
^{2}\rightarrow \infty $ has been used and so the form of 
(\ref{liouvdecohmaster1}) is valid for large times after the scattering of the
matter string with the defect. Formally, in order to avoid unphysical
singularities as $t\rightarrow 0$, we may replace the $1/t$ term by 
$1/(a+t)$, with $a$ a positive constant, i.e. 
\begin{equation}\label{liouvdecohmaster}
\overset{.}{\partial _{t}\rho _{Matter}}=i\left[ \rho _{Matter},H\right] -\,%
\frac{\widetilde{\gamma }}{a+t}\left[ \overline{u}_{j},\left[ \overline{u}%
^{j},\rho _{Matter}\right] \right] ~.
\end{equation}
with $a>0$. But we stress again, that at short times after the collision our 
$\sigma $-model perturbation theory breaks down. It is also for this reason
that the above-derived master equation is considered as rather
``phenomenological''.

\subsection*{Vacuum Energy contribution to the master equation for Liouville
Decoherence}

The above-described effects is not the only contribution to decoherence. If
there is a vacuum energy in the space-time over which the non-critical
string propagates, then the above-described foam effects will also
contribute to a novel type of decoherence, associated with the vacuum energy.

To determine this effect we first notice that the renormalization-group
relevant deformations (\ref{path1}) require Liouville dressing in order to
restore the conformal invariance of the $\sigma$-model. There are two ways
one can proceed in this matter. As we shall demonstrate below, the two
approaches are physically equivalent, as far as the (perturbative)
calculation of the back reaction onto space time is concerned.

\noindent \textit{Method I.}

The first method concerns dressing of the boundary operator (\ref{path1}) 
\begin{eqnarray}  \label{boundary}
V_{L,\mathrm{boundary}} = \int _{\partial \Sigma} e^{\alpha_i \phi}
y_i(t)\partial_nX^i, \qquad \alpha_i = -\frac{Q}{2} + \sqrt{\frac{Q^2}{2} +
(1-h_i)}
\end{eqnarray}
where $h_i$ is the boundary conformal dimension, and $Q^2$ is the induced
central charge deficit on the boundary of the world-sheet. In what follows
we deal first with Euclidean target-spaces prior to Liouville dressing.

The rate of change of $Q^2$ with respect to world-sheet scale $\mathcal{T}
\equiv \mathrm{ln}|L/a|^2 \sim \epsilon^{-2}$ is given by means of
Zamolodchikov's c-theorem~\cite{zam}, and it is found to be of order~\cite%
{recoil} $\overline{u}_i^2~\epsilon^4 $, as being proportional to the square
of the renormalization-group $\beta^i$ functions ($i = u_i$): $\frac{%
\partial Q^2}{\partial \mathcal{T}} \propto -\beta^i \mathcal{G}_{ij}
\beta^j $, where $\mathcal{G}_{ij} = \frac{1}{\varepsilon^2}\delta _{ij} +
\dots$, is the Zamolodchikov metric in coupling constant $u$-space. This
implies that $Q^2(t) = Q_0^2 + \mathcal{O}(\epsilon^2)$, where $Q_0^2$ is
constant.

We shall distinguish two cases for $Q_0$. The first concerns the case where $%
Q_0 \ne 0$ (and by appropriate normalization may be assumed to be of order $%
\mathcal{O}(1)$). This is the case of strings living in a non-critical space
time dimension. The other pertains to the case where the only source of
non-criticality is the impulse deformation, i.e. $Q_0 =0$. In the former
case, one has a Liouville dimension $\alpha_i \sim \epsilon^2 $, while in
the latter $\alpha_i \sim \epsilon$.

We next rewrite the boundary operator (\ref{boundary}) as a bulk operator,
using Stokes' theorem, and then manipulate it as follows: 
\begin{eqnarray}
&~&V_{L,\mathrm{boundary}}=\int_{\Sigma }\partial _{\alpha }\left( e^{\alpha
_{i}\phi }y_{i}(t)\partial ^{\alpha }X^{i}\right) =  \notag
\label{boundary2} \\
&&~=\int_{\Sigma }\alpha _{i}e^{\alpha _{i}\phi }~y_{i}(t)\partial _{\alpha
}\phi \partial ^{\alpha }X^{i}+\int_{\Sigma }e^{\alpha _{i}\phi }{\dot{y}}%
_{i}(t)\partial _{\alpha }t\partial ^{\alpha }X^{i}+\int_{\Sigma }e^{\alpha
_{i}\phi }y_{i}(t)\partial ^{2}X^{i}=  \notag \\
&~&\int_{\Sigma }\alpha _{i}e^{\alpha _{i}\phi }~y_{i}(t)\partial _{\alpha
}\phi \partial ^{\alpha }X^{i}+\int_{\Sigma }e^{\alpha _{i}\phi }\partial
_{\alpha }\left( y:i(t)\partial ^{\alpha }X^{i}\right)
\end{eqnarray}%
The first term in the last line describes an off-diagonal metric
contribution (in our chosen coordinate system) of the form (\ref{recmetr}): 
\begin{equation}
g_{0i}=\alpha _{i}y_{i}(t)  \label{eqbound}
\end{equation}

\noindent \textit{Method II.}

In the second method~\cite{recoil}, one rewrites the boundary operator (\ref%
{path1}) as a bulk total world-sheet derivative operator, and then
Liouville-dresses the bulk operator i.e. 
\begin{eqnarray}  \label{bulk}
V_{L, \mathrm{bulk}} = \int _{\Sigma} e^{\alpha_i \phi} \partial_\alpha
\left(y_i(t)\partial^\alpha X^i \right), \qquad \alpha_i = -\frac{Q}{2} + 
\sqrt{\frac{Q^2}{2} + (2-\Delta_i)}
\end{eqnarray}
where $\Delta_i$ is the conformal dimension of the bulk operator. The
central charge deficit $Q$ is of the same order $Q^2 = Q_0^2 + \mathcal{O}%
(\epsilon^2)$ as in the boundary case, which implies again that $\alpha_i
\sim \epsilon^2$ if $Q_0 \ne 0$, and $\alpha_i \sim \epsilon$ if $Q_0 = 0$.

For the bulk operator (\ref{bulk}) one has: 
\begin{eqnarray}
&~&V_{L,\mathrm{bulk}}=\int_{\Sigma }\partial _{\alpha }\left( e^{\alpha
_{i}\phi }y_{i}(t)\partial ^{\alpha }X^{i}\right) -\int_{\Sigma }\alpha
_{i}e^{\alpha _{i}\phi }~y_{i}(t)\partial _{\alpha }\phi \partial ^{\alpha
}X^{i}=  \notag  \label{bulk2} \\
&&~=\int_{\partial \Sigma }e^{\alpha _{i}\phi }y_{i}(t)\partial
_{n}X^{i}-\int_{\Sigma }\alpha _{i}e^{\alpha _{i}\phi }~y_{i}(t)\partial
_{\alpha }\phi \partial ^{\alpha }X^{i}
\end{eqnarray}%
From the second term of the last line of (\ref{bulk2}) one obtains an
induced target-space metric contribution 
\begin{equation}
g_{0i}=-\alpha _{i}y_{i}(t)  \label{eqbulk}
\end{equation}%
which differs from the induced metric (\ref{eqbound}) by an overall minus
sign. The latter is innocuous, and can be absorbed in a rescaling of the
coordinates. Moreover, since in the respective master equations (\ref%
{master4}) the metric appears quadratically, the sign is irrelevant in this
respect.

We now mention that~\cite{kmw} the logarithmic algebra implies a non-trivial
infrared fixed point, which in the case $Q_0 \ne 0$ is determined by $\phi_0
= \epsilon^{-2} \sim \mathrm{ln}(L/a)^2 \to \infty$, where $\phi_0$ is the
Liouville field world-sheet zero mode. Thus, $\alpha_i \phi_0 $ is finite as 
$\epsilon \to 0^+$. Therefore, as expected from the restoration of the
conformal invariance by means of the Liouville dressing, one can now take
safely the infra-red limit $\epsilon \to 0^+$ in the above expressions. It
is then easy to see that one is left \textit{in both cases} with the
target-space metric (\ref{recoil}), thereby proving the equivalence of both
approaches at the infrared fixed point.

In the case $Q_0 =0$, the running central charge deficit $Q^2 =\mathcal{O}%
(\epsilon^2)$. Recalling~\cite{ddk} that the above formul\ae\, imply a
rescaling of the Liouville mode by $Q \sim \epsilon$, so as to have a
canonical kinetic $\sigma$-model term~\footnote{%
Notice that this rescaling becomes a trivial one in the case where $Q_0 \ne
0 $.}, and that in this case it is the $\phi_0/Q$ which is identified with $%
\mathrm{ln}(L/a)^2 \sim \epsilon^{-2}$ as pertaining to the covariant
world-sheet cutoff, one observes that again $\alpha_i~\phi$ is finite as $%
\epsilon \to 0^+$, and hence similar conclusions are reached concerning the
equivalence of the two methods of Liouville dressing of the impulse operator
(\ref{path1}).

The recoil-induced metric (\ref{eqbound}) (or, equivalently (\ref{eqbulk})),
implies novel decoherence contributions to the master equation. As discussed
in the text, if there are such contributions, then they show up as
entanglement contributions for the case that the string propagates in a
de-Sitter background. In such a case the space-time is not Ricci flat, since
the corresponding Ricci tensor reads 
\begin{equation}
R_{\mu \nu }=\Omega g_{\mu \nu }  \label{ricci}
\end{equation}%
where $\Omega $ denotes vacuum energy, which may even be allowed to depend
on cosmic time. From a $\sigma $-model view point, such backgrounds are not
conformal, since the corresponding graviton $\beta $-function is precisely
given by the Ricci tensor to leading order in the Regge-slope ($\alpha
^{\prime }$) perturbative expansion.From the master equation (\ref{master}),
then, one obtains in this case the specific master equation (\ref{master4})
in the text.

Combining the two types of effects, D-particle foam (\ref{liouvdecohmaster})
and Vacuum energy contributions (\ref{ricci}), one may arrive at the
following master equation to be used in the text: 
\begin{eqnarray}
\overset{.}{\partial _{t}\rho _{Matter}}&=&i\left[ \rho _{Matter},H\right]
-\,\Omega_{\mathrm{total}}\left[ \overline{u}_{j},\left[ \overline{u}%
^{j},\rho _{Matter}\right] \right]~.  \label{liouvdecohmaster2}
\end{eqnarray}
where 
\begin{equation}  \label{omegatotal}
\Omega_{\mathrm{total}} (t) = \Omega_0 + \frac{\widetilde \gamma}{a + t} + 
\frac{\widetilde{\Gamma}}{1 + bt^2}
\end{equation}
with the various forms having been defined in the text.

The constant $\Omega_0 > 0$ has been added to take possible account of other
types of foam, or cosmological constant contributions. It is only this
constant (positive) part that is responsible for exponential damping factors
in physical quantities such as oscillation probabilities.

\subsection*{Sub-leading Ohmic-type Contributions of the D-particle Foam to
the master equation}

Before closing this discussion we would like to comment briefly on the form
of subleading effects (as $\varepsilon \to 0$) associated with the operators
pertaining to the D-particle coordinates $y_i$ in (\ref{path}). Note that
the relevant world-sheet operator associated with this is $\varepsilon
\Theta_\varepsilon (t)$.

Such terms would contribute the following environmental entanglement in the
master equation (\ref{master}): 
\begin{eqnarray}  \label{entangl2}
&~&\mathrm{Sub-leading~Liouville~Entanglement~in~Eq.~(\ref{master})} = 
\notag \\
&~& :\beta^{y_i}\mathcal{G}_{y_iy_j}[y^j, \rho] + :\beta^{y_i}\mathcal{G}%
_{y_iu_j}[u^j, \rho]:
\end{eqnarray}
Such terms are additional to the leading order terms exhibited in (\ref%
{liouvdecohmaster}) above. The components of the Zamolodchikov's metric in $%
(u_i,~y^i)$ ``space'' can be found by means of explicit computation using
the (logarithmic) conformal algebra of the pertinent deformations~\cite{kmw}%
. To leading order in $\varepsilon \to 0$ they read: 
\begin{equation}
\mathcal{G}_{y_iy_j} \sim -\varepsilon ^2 \delta_{ij}~, \qquad \mathcal{G}%
_{y_iu_j} \sim \delta_{ij}  \label{corrfunct}
\end{equation}
where we note the relative minus sign of the first correlator. The above
expressions refer to Euclidean formalism, with $\varepsilon ^2 > 0$. On
using (\ref{betafnct}) and passing into exactly marginal couplings $%
\overline y_i \equiv \frac{y_i}{\varepsilon}$, $\overline u_j \equiv \frac{%
u_j}{\varepsilon}$, with $i,j = 1, \dots d$ spatial indices, we obtain for
the entanglement terms (\ref{entangl2}) (assuming antisymmetric quantum
operator ordering throughout): 
\begin{eqnarray}  \label{entangl3}
\mathrm{Sub-leading~Liouville~Entanglement~in~Eq.~(\ref{master})} = \frac{%
\varepsilon^6}{2} \overline{y}_i\overline{y}^i - \frac{\varepsilon^4}{2} 
\overline{y}_i\overline{u}^i~.
\end{eqnarray}
Taking into account that $\varepsilon ^{-2} \sim t$ is the (Minkowski)
target time, and using (\ref{trnsf}), we may rewrite the entanglement (\ref%
{entangl3}) as follows (again formally regularizing for short times, where
our approach is not valid): 
\begin{eqnarray}  \label{entangl4}
&~&\mathrm{Sub-leading~Liouville~Entanglement~in~Eq.~(\ref{master})} = 
\notag \\
&~&\frac{1}{2(a_1 + t^3)} [\overline{y}_i,~[\overline{y}^i, \rho ]] - \frac{1%
}{2M_P( a_2 + t^2)} r [\overline{y}_i,~[\widehat{p}^i, \rho]]~.
\end{eqnarray}
where $a_{1,2}$ are appropriate positive cut-off constants, serving as
regulators in the $t \to 0$ limit. Upon noting that the D-particle
coordinate operator $\overline{y}^i $ can be identified with the (stringy)
probe coordinate operator $\widehat{x}_i$ in the entangled state, one
observes that the first term in (\ref{entangl4}) acquires a conventional
ohmic form $[\widehat x,~[\widehat x,~\rho]]$ of the type considered in (\ref%
{master5}), with a time-dependent coefficient given by (\ref{entangl4})
above. This motivates the amalgamated form of the master equation considered
in this work, combining ohmic and D-particle foam effects.

The randomness assumption (\ref{random}) eliminates the second term of (\ref%
{entangl4}), which would otherwise constitute an additional entanglement
term in the master equation, of a type also encountered in customary
decoherent systems. However, this term may be present in models of foam in
which there is an ordered bulk current of D-particle defects crossing the D3
brane world, which could contribute an average recoil velocity $\langle
u_{i}\rangle \neq 0$ on the three-dimensional brane world. The latter type
of models would result in modified dispersion relations for low-energy
probes, but these need to satisfy severe phenomenological constraints~\cite%
{jacobson}; however, it has been argued\cite{sakharov} that only photons,
and probably gauge bosons, could exhibit entanglement with the D-particle
foam, for purely stringy reasons that we shall not discuss in this work.
These latter types of probes suffer less severe phenomenological
restrictions at present.

This completes our semi-rigorous (rather phenomenological) analysis of
D-particle foam effects and their r\^ole in inducing decoherence of quantum
matter propagating in it.

\end{document}